\documentclass[12pt,epsf,rotate]{article}
\usepackage[dvips]{graphicx}
\usepackage{color}
\usepackage{amsfonts,amssymb,amsmath}
\usepackage{graphics}
\usepackage{graphicx}
\input epsf
\usepackage[figuresright]{rotating}
\def\be{\begin{equation}}
\def\ee{\end{equation}}
\def\ber{\begin{eqnarray}}
\def\eer{\end{eqnarray}}
\def\dint{\mathop{\intop\kern-0.5em\intop}}
\def\ovc#1{\displaystyle\mathop{#1}^{\kern0.2em\circ}}

\begin{document}

\begin{center}
{\bf\Large On ``soft" physics at the CERN\\[1ex] Large Hadron Collider}\footnote{This
article is an extended version of the talk prepared for the
International Conference on High Energy
physics ICHEP2010, July 21-28, 2010, Paris, France.} \\

\vspace{4mm}

{\bf\large Andrei A. Arkhipov}\\[1ex]
{\it State Research Center ``Institute for High Energy Physics" \\
 142280 Protvino, Moscow Region, Russia}\\
\end{center}

\begin{abstract}
Three tightly inter-related topics have been discussed: the $pp(p\bar
p)$ total cross section; the single diffraction dissociation cross
section; the $p(\bar p)d$ total cross section and the defect of the
total cross section in scattering from deuteron.
\end{abstract}

{\bf Keywords:} Froissart theorem, total cross section, single
diffraction dissociation, three-body forces

{\bf PACS:} 11.55.-m, 11.80.-m, 13.85.-t

\vspace{1cm}

\rightline{\large\it ``To reach great heights,}
\vspace{2mm}\rightline{\large\it one must possess great depth."}
\section{Introduction: Global Features of the Total Cross Sections}
%\vspace{1cm}
\noindent

I would like to discuss here some aspects of the so called ``soft"
physics, which is known as the physics of long-range strong
interactions. More precisely, I will concentrate on three deeply
inter-related topics where I have personal contribution to
theoretical basis of our fundamental understanding the hadronic
interactions:
\begin{itemize}
\item the $pp(p\bar p)$ total cross section
\item the single diffraction dissociation cross section
\item the $p(\bar p)d$ total cross section and the defect of the total
cross section in scattering from nuclei.
\end{itemize}
Most of the material that I present here is taken from my works over
the years. Some details may be found in recent article \cite{1} and
references therein.

The $pp$ total cross section is probably one that will measured at
the LHC in the first. Fig. 1 shows the sets of data points for the
$pp$ and $p\bar p$ total cross sections measured up to now.

\begin{figure}
\begin{picture}(640,205)
\put(-10,10){\scalebox{.7}{\includegraphics[]{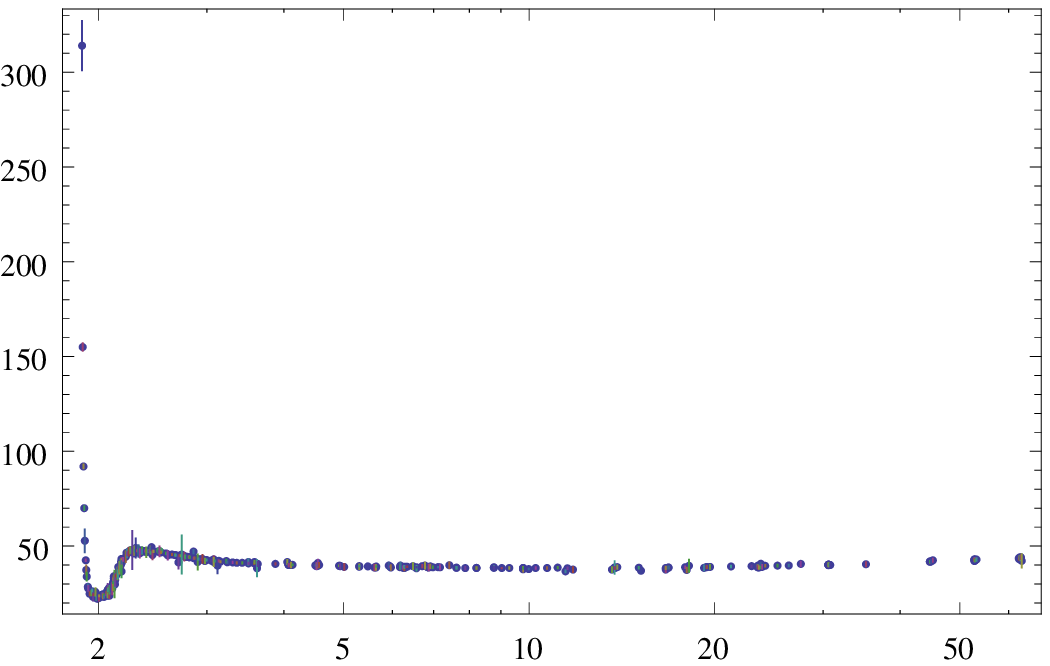}}}
\put(90,0){$\sqrt{s}\, (\rm GeV)$}
\put(-25,65){\rotatebox{90}{$\sigma_{tot}^{pp} (\rm mb)$}}
\put(185,125 ){A}
\put(230,10){\scalebox{.7}{\includegraphics[]{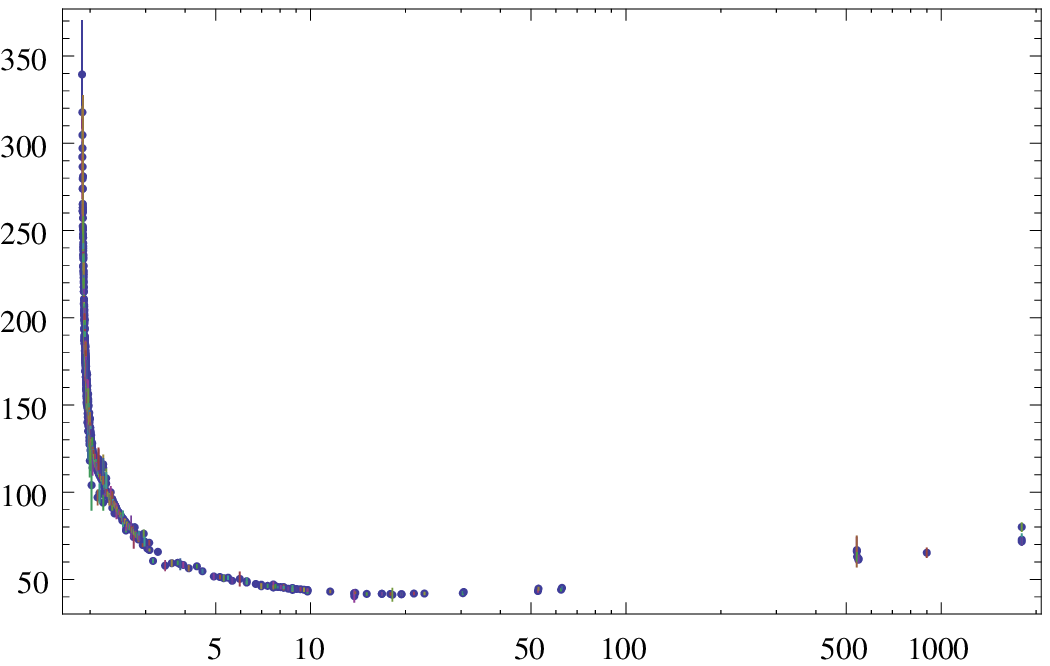}}}
\put(320,0){$\sqrt{s}\, (\rm GeV)$}
\put(210,65){\rotatebox{90}{$\sigma_{tot}^{p\bar p} (\rm mb)$}}
\put(425,125){B}
\end{picture}

\begin{picture}(640,205)
\put(-10,10){\scalebox{.7}{\includegraphics[]{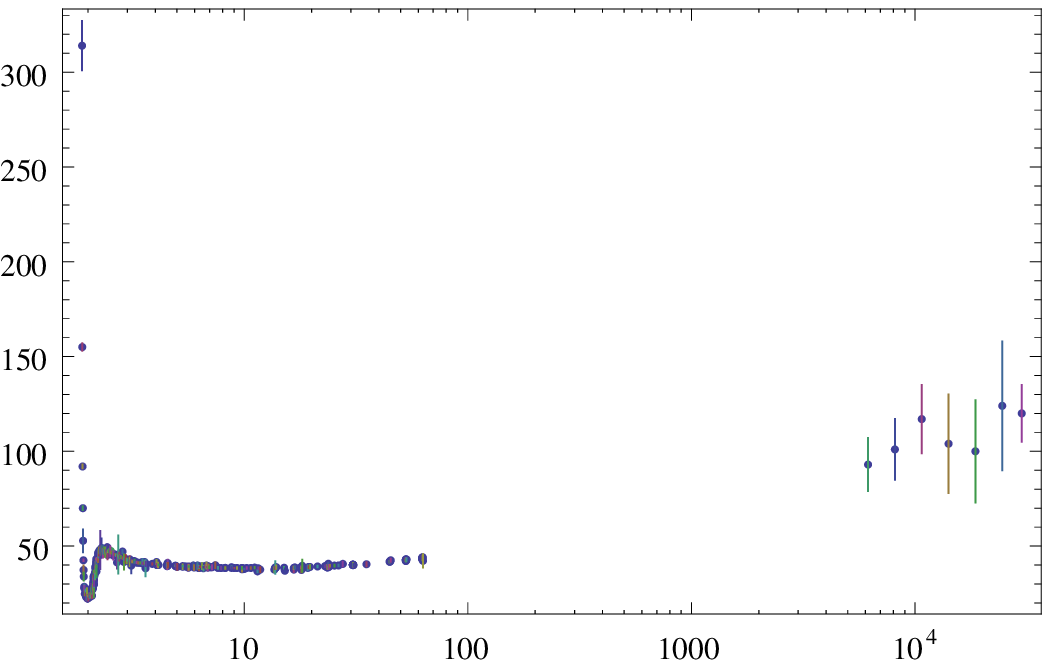}}}
\put(80,0){$\sqrt{s}\, (\rm GeV)$}
\put(-25,65){\rotatebox{90}{$\sigma_{tot}^{pp} (\rm mb)$}}
\put(185,125){C}
\put(230,10){\scalebox{.7}{\includegraphics[]{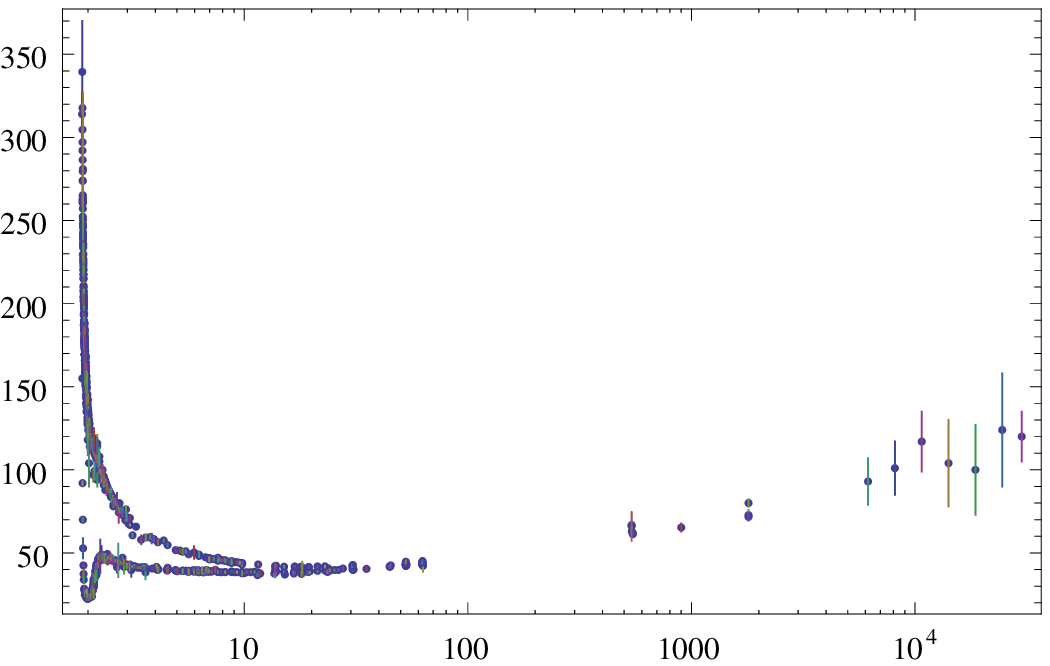}}}
\put(320,0){$\sqrt{s}\, (\rm GeV)$}
\put(210,65){\rotatebox{90}{$\sigma_{tot}^{pp(\bar p)} (\rm mb)$}}
\put(425,125){D}
\end{picture}
\vspace{1cm} \caption{The world data set on the $pp$ and  $p\bar p$
total cross sections: the experimental (accelerator) data (270
points) on the proton-proton total cross-section (A); the
experimental (accelerator) data (444 points) on the proton-antiproton
total cross-section (B); the full experimental data (277 points) on
the proton-proton total cross-section including the cosmic ray data
(C); combined experimental data on the $pp$ and $p\bar p$ total cross
sections (D). All presented data extracted from
\cite{2}.}\label{fig1}
\end{figure}

Some time ago I have investigated whether a unified formula, similar
to the Planck formula for black body radiation, can be obtained which
simultaneously described hadron total cross sections in the whole
range of energies from the most low energy to the most high one. It
is remarkable that such formula can really be attainable, and here, I
am going to show the results of our studies. But, first of all, let
me say a few words instead of introduction.

From Fig.~\ref{fig1} it follows that the $pp$ and $p\bar p$ total
cross sections rapidly increase at approaching the elastic scattering
threshold. The large cross section near to the elastic scattering
threshold was explained by the greatest physicist -- Nobel Prize
winner of the year 1967 Hans Albrecht Bethe without any additional
assumption but only through a straightforward application of
nonrelativistic quantum mechanics providing the cross sections of
just the right magnitude. Bethe has proved that the cross section for
the elastic scattering of slow nucleons is large due to the cross
section is inversely proportional to nucleon velocity \cite{3}
\begin{equation}\label{Bethe}
\sigma_{tot}(s) = \sigma_{el}(s) \sim \frac{\sigma_0}{v} =
\sigma_0\sqrt{\frac{s}{s-4m_p^2}}, \quad s\rightarrow 4m_p^2.
\end{equation}
Next, it has been revealed itself the dip structure in the
proton-proton total cross-section at $\sqrt{s}\simeq 2\,\mbox{GeV}$
compared to the proton-antiproton total cross-section where such
structure is absent. At last, the increase of the $pp$ total cross
section has been discovered at the CERN ISR \cite{4} and then the
effect of rising $p\bar p$ total cross sections was confirmed at the
Fermilab accelerator \cite{5} and CERN $Sp\bar p S$ \cite{6}.

Although nowadays we have in the framework of local quantum field
theory a gauge model of strong interactions formulated in terms of
the known QCD Lagrangian, its relations to the ``soft" hadronic
physics are far from desired. In spite of almost 40 years after the
formulation of QCD we still cannot obtain from the QCD Lagrangian the
answer to the question why all the hadronic total cross-sections grow
with energy. We cannot predict total cross-sections in an absolute
way starting from the fundamental QCD Lagrangian as well mainly
because the effective QCD running coupling is not small and thus we
cannot use perturbation theory. It is well known, e.g., that
nonperturbative contributions to the gluon propagator influence the
behaviour of ``soft" hadronic processes and the knowledge of the
infrared behaviour of QCD is certainly needed to describe the ``soft"
hadronic physics in the framework of QCD. Unfortunately, today we
don't know the whole picture of the infrared behaviour of QCD, we
have some fragments of this picture though (see e.g. Ref. \cite{7}).
The understanding of the ``soft" physics is of high interest because
it has an intrinsically fundamental nature.

The behaviour of hadronic total cross sections at high energies is a
wide and much discussed topic in high-energy physics community; see
e.g. the proceedings of famous Blois Workshops. At present time there
are a lot of different models which provide different energy
dependencies of hadronic total cross sections at high energies and
different predictions for a range of the LHC energies.

All different phenomenological models can conditionally be separated
into two groups in according to two forms of strong interaction
dynamics used: t-channel form and s-channel one. The fundamental
quantity in the t-channel form of strong interaction dynamics is some
set of Regge trajectories:
\[
t-\mbox{channel form}\qquad \Longleftrightarrow \qquad \alpha_R(t),
\]
where subscript $R$ enumerates different Regge trajectories which are
the poles in the t-channel partial wave amplitudes for the given
process. The first group contains the Regge-type models with
power-like, $s^{\alpha_P(0)-1}$, behaviour of hadronic total cross
sections. Here $\alpha_P(0)$ is an intercept of the supercritical
Pomeron trajectory: $\alpha_P(0)-1=\Delta<<1$,
$\Delta>0\,(\Delta=0.0808)$ is responsible for the growth of hadronic
cross sections with energy. There are a lot of people who works with
such a type of Regge-pole models; see excellent review articles and
numerous conferences talks presented by Prof. Peter Landshoff from
Cambridge (England) \cite{8} and references therein.

Some part of scientific community works in the field of s-channel
form of strong interaction dynamics. The fundamental  quantity here
is an effective interaction radius of fundamental forces:
\[
s-\mbox{channel form} \qquad \Longleftrightarrow \qquad
R_{\alpha}(s),
\]
where subscript $\alpha$ enumerates different types of hadrons and
fundamental forces acting between them. The s-channel form of
dynamics allows one to create a physically transparent and visual
geometric picture of strong interactions for hadrons. It should be
mentioned here the founders of geometric (impact) picture for strong
interactions: the great theoretical physicist -- Nobel Prize winner
for the year 2005 Roy Glauber \cite{9} and the greatest theoretical
physicist -- Nobel Prize winner for the year 1957 C.N. Yang
\cite{10}. I'd like to emphasize the attractive features of this form
of strong interaction dynamics:
\begin{itemize}
\item universality (existence of pion with $m_{\pi}\not=0$)
\[
\fbox{$\displaystyle R_{\alpha}(s) \sim \frac{r_{\alpha}}{m_{\pi}}\ln
\frac{s}{s_0},\quad s \rightarrow \infty$}
\]
\item compatibility with the general principles of relativistic
quantum theory.
\end{itemize}
My personal preference is in favour of  the s-channel form of strong
interaction dynamics. This is, first of all, related to the fact that
the Regge phenomenology with the super-critical Pomeron exchange
breaks down the fundamental principles of relativistic quantum theory
such as unitarity, and this fact is often overlooked. In our opinion
only this pathology of the super-critical Pomeron model is enough to
reject the model from consideration. Moreover, accurate and complete
analysis of experimental data on hadron total cross sections has
shown that the super-critical Pomeron model is disfavoured from
statistical point of view \cite{11,12}, and experimental results from
HERA \cite{13} lead us to the same conclusion: The soft Pomeron
phenomenology developed cannot incorporate the HERA data on structure
function $F_2$ at small $x$ and data on total $\gamma^{*}p$ cross
section from $F_2$ measurements as a function of $W^2$ for different
$Q^2$. At last, Regge phenomenology without serious modifications
cannot be applied to the description of experimental data on single
diffractive dissociation cross sections in $p\bar p$ collisions; see
e.g. discussion in \cite{14}.
\begin{figure}
\begin{picture}(640,205)
\put(-10,10){\scalebox{.7}{\includegraphics[]{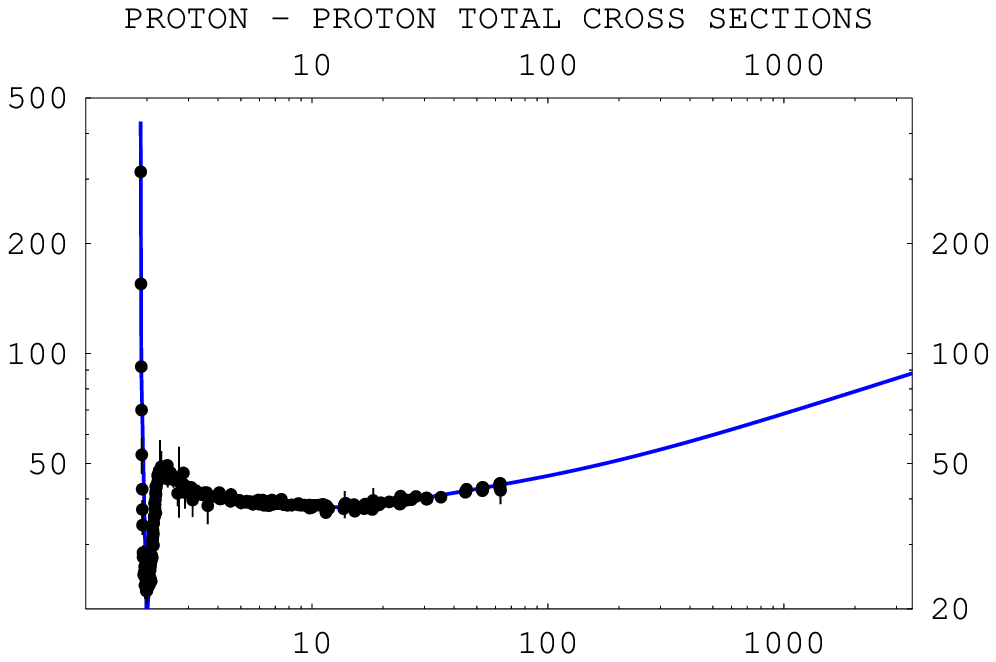}}}
\put(80,0){$\sqrt{s}\, (\rm GeV)$}
\put(-25,65){\rotatebox{90}{$\sigma_{tot}^{pp} (\rm mb)$}}
\put(160,110 ){A}
\put(230,10){\scalebox{.7}{\includegraphics[]{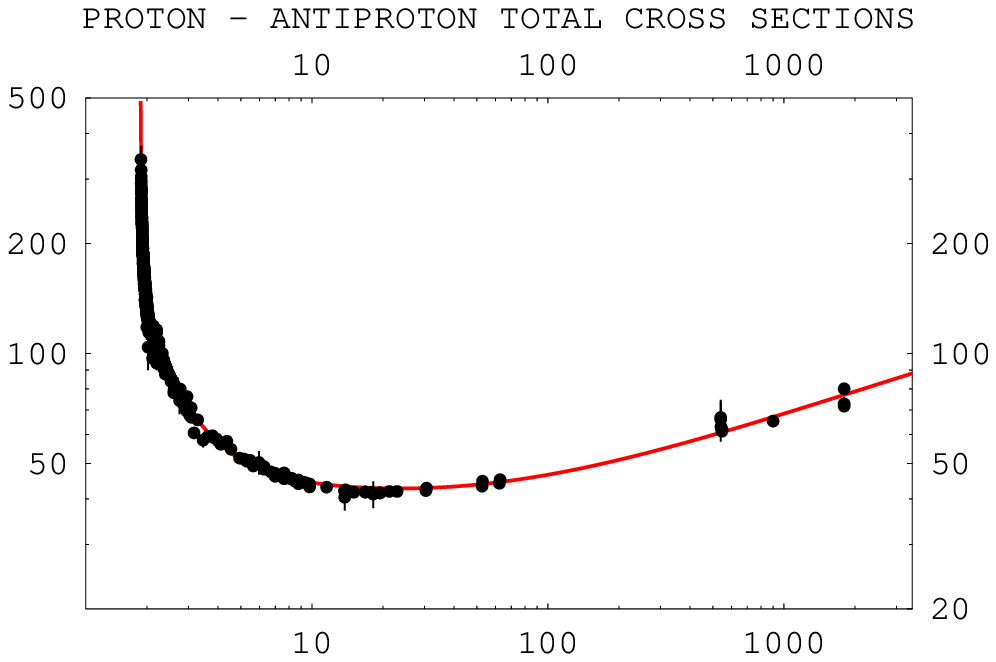}}}
\put(320,0){$\sqrt{s}\, (\rm GeV)$}
\put(210,65){\rotatebox{90}{$\sigma_{tot}^{p\bar p} (\rm mb)$}}
\put(400,110){B}
\end{picture}

\begin{picture}(640,205)
\put(-10,10){\scalebox{.7}{\includegraphics[]{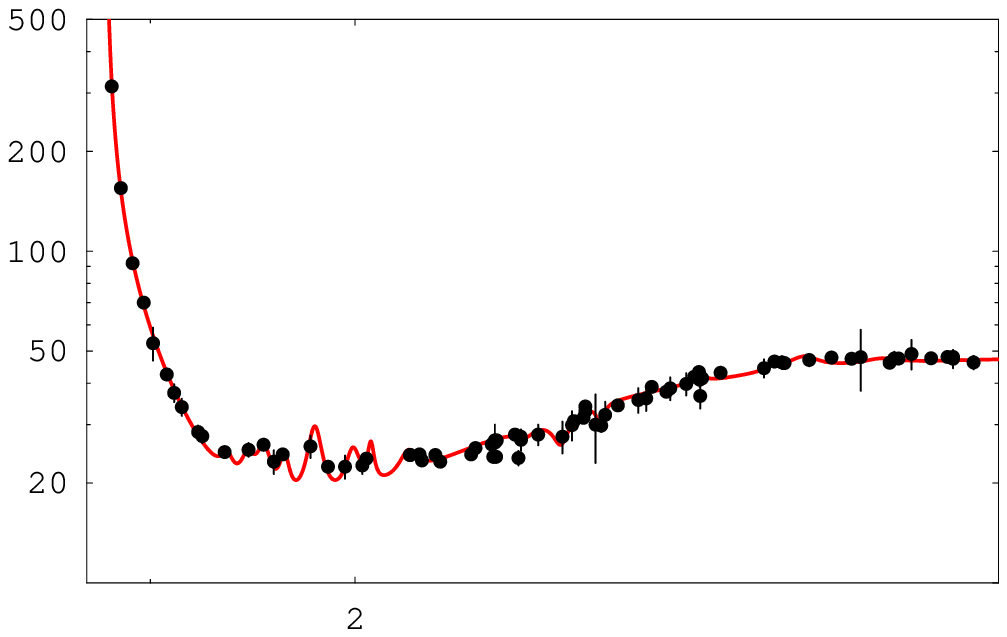}}}
\put(80,0){$\sqrt{s}\, (\rm GeV)$}
\put(-25,65){\rotatebox{90}{$\sigma_{tot}^{pp} (\rm mb)$}}
\put(180,125){C}
\put(230,10){\scalebox{.7}{\includegraphics[]{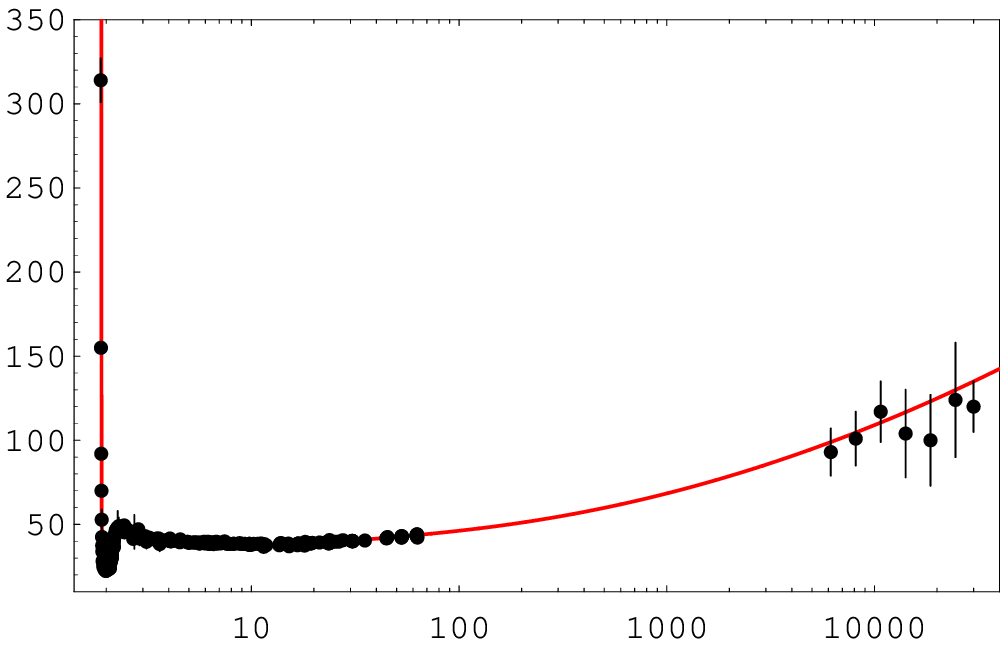}}}
\put(320,0){$\sqrt{s}\, (\rm GeV)$}
\put(210,65){\rotatebox{90}{$\sigma_{tot}^{pp} (\rm mb)$}}
\put(420,125){D}
\end{picture}
\vspace{1cm} \caption{The global description of the $pp$ and  $p\bar
p$ total cross sections: the full experimental (accelerator) data on
the proton-proton total cross-section (A); the full experimental data
on the proton-antiproton total cross-section (B); the proton-proton
total cross-section at low energies (C); the full experimental data
on proton-proton total cross-section including the cosmic ray data
(D).}\label{fig2}
\end{figure}

\section{Global Description of the $pp$ and $p\bar p$ Total Cross Sections}

As mentioned above, we have tried to derive a possibly simple
theoretical formula which would describe the global structure of $pp$
and $p\bar p$ total cross-sections in the whole range of energies
available up today making such derivation in a maximally model
independent way using general structures and general theorems in
Quantum Field Theory. The fit to the experimental data with the
formula was made, and it was shown that there is a very good
correspondence of the theoretical formula to the existing
experimental data obtained at the accelerators. Moreover, it turned
out that there is a very good correspondence of the formula to all
existing cosmic ray experimental data  as well: The predicted values
for $\sigma_{tot}^{pp}$ obtained from description of all existing
accelerators data are completely compatible with the values obtained
from cosmic ray experiments \cite{15}. The global description of the
proton-proton and antiproton-proton total cross sections is shown in
Fig.~\ref{fig2}.

The theoretical formula describing the global properties of
proton-proton and proton-antiproton total cross sections has the
following quite a simple form \cite{16}:
\begin{equation}
\sigma_{tot}(s) = [1+\chi(s)]\sigma^{\,a}_{tot}(s).\label{sigmatot}
\end{equation}
That structure has been appeared as consistency condition to fulfil
the unitarity requirements in two-particle and three-particle sectors
simultaneously. In according to this structure the total cross
section is represented in a factorized form. The first factor is
responsible for the behavior of total cross section at low energies
with the universal energy dependence at elastic threshold, it has a
complicated resonance structure, and $\chi(s)$ tends to zero at
$s\rightarrow \infty$. The other factor describes high energy
asymptotic of total cross section, and it has the universal energy
dependence predicted by the general theorems in Quantum Field Theory.
For this factor one obtains
\begin{eqnarray}
\sigma^{\,a}_{tot}(s) &=& 2\pi B_{el}(s) + 4\pi (1-\beta)B_{sd}(s) =
2\pi B_{el}(s)[1 + 2\gamma(1-\beta)]\nonumber\\[1ex]
&=& \pi R_2^2(s) + 2\pi
(1-\beta)R_0^2(s)\, = \,\pi R_2^2(s)[1 + 2\gamma(1-\beta)],
\label{asigma}
\end{eqnarray}
where $B_{el}(s)$ is the slope of diffraction cone in forward elastic
scattering processes
\[
B_{el}(s) =
\Big[\frac{d}{dt}\ln\big(\frac{d\sigma_{el}}{dt}(s,t)\big)\Big]_{t=0},
\]
$R_2(s)$ -- is the effective radius of two-nucleon forces related to
the slope $B_{el}(s)$ of diffraction cone by Equality $B_{el}(s) =
R_2^2(s)/2$, $B_{sd}(s)$ is the slope of diffraction cone for
inclusive diffraction dissociation processes at a special value of
missing mass
\[
B_{sd}(s) =
\Big[\frac{d}{dt}\ln\big(\frac{d\sigma_{sd}}{dt}(s,t,M_X^2)\big)|_{M_X^2
= 2m_p^2}\Big]_{t=0},
\]
$R_0(s)$ is the effective radius of three-nucleon forces related to
the slope $B_{sd}(s)$ in the same way $B_{sd}(s) = R_3^2(s)/2$ as the
effective radius of two-nucleon forces is related to the slope
$B_{el}(s)$ of diffraction cone in elastic scattering processes,
$\beta$ is slowly energy dependent dimensionless quantity from
interval $0\leq \beta \leq 1/4$
\[
\beta = \frac{x^2_{inel}}{4(1+x^2_{inel})},\quad
x^2_{inel}=\frac{R_0^2(s)}{R_d^2}=\frac{2B_{sd}(s)}{R_d^2},
\]
$R_d$  characterizes the internucleon distance where a two-nucleon
bound state -- the deuteron has arising. $\beta$ tends to 1/4 at
$s\rightarrow \infty$ and $\beta \ll 1$ up to LHC energies, and
$\gamma = B_{sd}(s)/B_{el}(s) = R_0^2(s)/R_2^2(s)$ obviously. From
the Froissart bound it follows $\gamma < 2$.

\begin{figure}[]
\bigskip
\begin{center}
\begin{picture}(288,198)
\put(15,10){\epsfbox{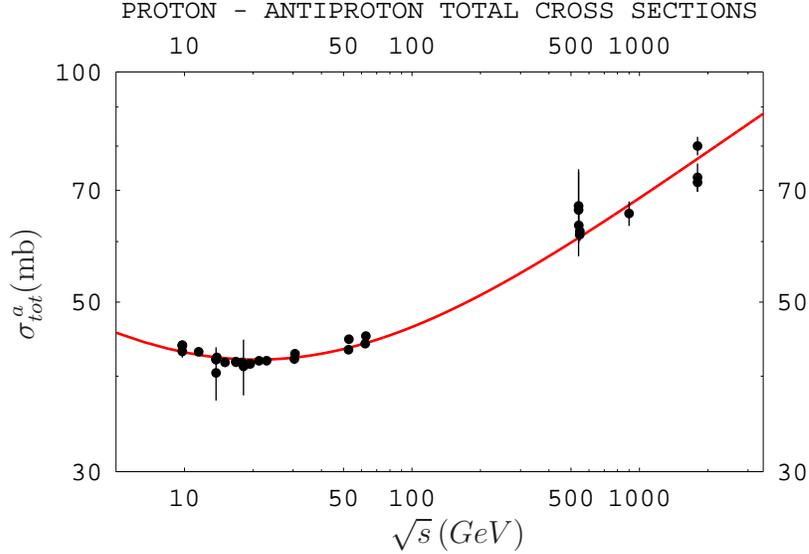}}
\put(144,0){$\sqrt{s}\, (GeV)$}
\put(0,77){\rotatebox{90}{$\sigma_{tot}^{\,a} (\rm mb)$}}
\end{picture}
\end{center}
\caption{The total proton-antiproton cross sections versus $\sqrt{s}$
compared with formula (\ref{fitsigma}). Solid line represents our fit
to the data.}\label{fig3}
\end{figure}

\begin{figure}[]
\begin{center}
\begin{picture}(288,188)
\put(15,10){\epsfbox{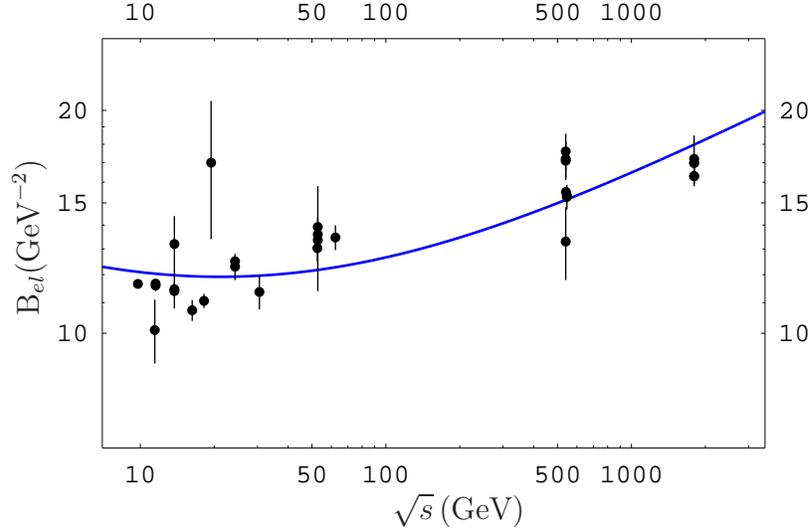}} \put(144,0){$\sqrt{s}\,
(\mbox{GeV})$} \put(0,77){\rotatebox{90}{$\mbox{B}_{el}
(\mbox{GeV}^{-2})$}}
\end{picture}
\end{center}
\caption{Slope $\mbox{B}_{el}$ of diffraction cone in $p\bar p$
elastic scattering. Solid line represents our fit to the
data.}\label{fig4}
\end{figure}

Using formula (\ref{sigmatot}), we have made the global fit \cite{16}
to the experimental data on $pp$ and $p\bar p$ total cross sections
shown above. The fitting procedure has been performed applying the
following experimentally established and theoretically justified
parameterizations
\begin{equation}\label{el-sd}
\frac{d\sigma_{el}}{dt}(s,t) =
\frac{d\sigma_{el}}{dt}(s,0)\exp[B_{el}(s)t],\quad
\frac{2s}{\pi}\frac{d\sigma_{sd}}{dtdM_X^2}(s,t,M_X^2) =
A(s,M_X^2)\exp[b(s,M_X^2)t].
\end{equation}
At the first step, we have made a weighted fit to the experimental
data on the proton-antiproton total cross sections in the range
$\sqrt{s}>10\, GeV$. The data were fitted with the function of the
form predicted by the Froissart bound in the spirit of our approach.
As mentioned above, a careful analysis of the experimental data and
comparative study of the known characteristic parameterizations have
revealed that statistically a ``Froissart-like" type parameterization
for proton-proton and proton-antiproton total cross sections is
strongly favoured \cite{11,12}. So
\begin{equation}
\sigma_{tot}^{\,a} = a_0 + a_2 \ln^2(\sqrt{s}/\sqrt{s_0})
\label{fitsigma}
\end{equation}
where $a_0, a_2, \sqrt{s_0}$ are free parameters. We accounted for
the experimental errors $\delta x_i$ (statistical and systematic
errors added in quadrature) by fitting to the experimental points
with the weight $w_i=1/(\delta x_i)^2$. Our fit yielded
\[
a_0 = (42.0479\pm 0.1086)\mbox{mb},\quad a_2 = (1.7548\pm
0.0828)\mbox{mb},
\]
\[
\sqrt{s_0} = (20.74\pm 1.21)\mbox{GeV}.
\]
The fit result is shown in Fig.~\ref{fig3} .
After that we have made a
weighted fit to the experimental data on the slope of diffraction
cone in elastic $p\bar p$ scattering. The experimental points and the
references, where they have been extracted from, are listed in
\cite{17}. The fitted function of the form
\[
B_{el} = b_0 + b_2 \ln^2(\sqrt{s}/20.74),
\]
which is also suggested by the asymptotic theorems of local quantum
field theory, has been used. The value $\sqrt{s_0}$ was fixed at
$\sqrt{s_0} = 20.74\,\mbox{GeV}$ from previous fit to the $p\bar p$
total cross sections data. Our fit yielded
\[
b_0 = (11.92\pm 0.15)\mbox{GeV}^{-2} ,\quad b_2 = (0.3036\pm
0.0185)\mbox{GeV}^{-2}.
\]
The fitting curve is shown in Fig.~\ref{fig4}.
\noindent From these fits in accordance with formula (\ref{asigma})
one obtains
\[
R^2_0(s)|_{\beta<<1} =
\left[5.267+0.4137\ln^2\sqrt{s}/20.74\right](GeV^{-2}).
\]
At the final stage to build a global (weighted) fit to the all data
on proton-antiproton total cross sections we have to choose an
appropriate parameterization for the function $\chi (s)$ in R.H.S. of
Eq. (\ref{sigmatot}). In fact, we have for the function $\chi (s)$
the theoretical expression in the form
\[
\chi (s) = \frac{C}{\kappa (s)R_0^3(s)},
\]
where
\[ \kappa^4 (s) =
\frac{1}{2\pi}\int_a^b dx
\sqrt{(x^2-a^2)(b^2-x^2)[(a+b)^2-x^2]},\quad a = 2m_p,\quad b =
\sqrt{2s+m_p^2}-m_p.
\]
It can be proved that $\kappa (s)$ has the following asymptotics.
\[
\kappa (s)\sim \sqrt{s},\quad s\rightarrow \infty; \qquad \kappa
(s)\sim \sqrt{s-4m_p^2},\quad s\rightarrow 4m_p^2.
\]
Here, among other things, we have reproduced from the first
principles the above mentioned Bethe's result. On the other side, at
high energy a Regge type asymptotic corresponding to secondary
Reggeons exchange has been arisen from the first principles as well.

For simplicity, the global fit to the all data on proton-antiproton
total cross sections was made with the function $\chi_{p\bar p}(s)$
of the form
\begin{equation}
\chi_{p\bar p}(s) =  \frac{c}{\sqrt{s-4m^2_p}R^3_0(s)}
\left(1+\frac{d_1}{\sqrt{s}} + \frac{d_2}{s} +
\frac{d_3}{s^{3/2}}\right),\label{chipap}
\end{equation}
where $m_p$ is the proton mass, and $c, d_1, d_2, d_3$ are free
parameters. The function $\chi_{p\bar p}(s)$ in Eq. (\ref{chipap})
contained the preasymptotic terms as well needed to describe the
region of middle (intermediate) energies -- this is a price that we
pay for simplicity. Our fit yielded
\[
d_1 = (-12.12\pm 1.023)\mbox{GeV},\quad d_2 = (89.98\pm
15.67)\mbox{GeV}^2,
\]
\[
d_3 = (-110.51\pm 21.60)\mbox{GeV}^3,\quad c = (6.655\pm
1.834)\mbox{GeV}^{-2}.
\]

As seen, the experimental data on proton-proton total cross sections
display a more complex structure at low energies than the
proton-antiproton ones. We can describe this complex structure
keeping the quantity $\sigma_{tot}^a(s)$ unchanged in Eq.
(\ref{sigmatot}) and taking  the following expression for
$\chi_{pp}(s)$
\begin{equation}
\chi_{pp}(s) = \left(\frac{c_1}{\sqrt{s-4m^2_N}R^3_0(s)} -
\frac{c_2}{\sqrt{s-s_{thr}}R^3_0(s)}\right)\left(1 + d(s)\right) +
\mbox{Res}(s),\label{chipp}
\end{equation}
\[
d(s) = \sum_{k=1}^{8}\frac{d_k}{s^{k/2}},\qquad \mbox{Res}(s) =
\sum_{i=1}^{N}\frac{C_R^i s_R^i
{\Gamma_R^i}^2}{\sqrt{s(s-4m_N^2)}[(s-s_R^i)^2+s_R^i{\Gamma_R^i}^2]},
\]
Our fit yielded
\[
c_1 = (192.85\pm 1.68)\mbox{GeV}^{-2},\quad c_2 = (186.02\pm
1.67)\mbox{GeV}^{-2},
\]
\begin{equation}
\fbox{$\displaystyle s_{thr}=(3.5283\pm
0.0052)\mbox{GeV}^2$}\,.\label{thresh}
\end{equation}
For the numerical values of the parameters $d_i$ and $C_R^i$ see
original paper \cite{16} and \cite{18}.

It should especially be emphasized that the global description of the
proton-proton total cross-section revealed the new ``threshold"
$s_{thr}=3.5283\, \mbox{GeV}^2$, which is near the elastic scattering
threshold. The position of the new ``threshold" has been determined
by the fit with a high accuracy. Note, all available experimental
data on the proton-proton total cross-section lie above this
``threshold". One could imagine an appearance of the new ``threshold"
as a manifestation of a new unknown particle:
\[
\sqrt{s_{thr}} = 2 m_p + m_{\cal L},\qquad m_{\cal L} =
1.833\,\mbox{MeV}.
\]
This particle was named \cite{18} as $\cal L$-particle from the word
{\it lightest}. Of course, the natural questions have been arisen.
What is the physical nature and dynamical origin of $\cal
L$-particle? Could $\cal L$-particle be related to the experimentally
observed diproton resonances spectrum? ..., and all that. Some
discussing that questions of fundamental importance may be found in
our articles \cite{19,20}.

At the LHC we predict
\begin{equation}\label{predict}
\sigma^{pp}_{tot}(\sqrt{s}=14\,\mbox{TeV}) = 116.53 \pm
3.52\,\mbox{mb}.
\end{equation}
Theoretical uncertainty in (\ref{predict}) has been calculated by one
deviation in parameter $a_2$ in Eq. (\ref{fitsigma}). It should to be
compared to the best even though very crude estimate based on {\it
Pomeron Physics and QCD} \cite{8}
\begin{equation}\label{9}
\sigma^{LHC} = 125 \pm 35\,\mbox{mb},
\end{equation}
presented by Peter Landshoff at the Conference "Diffraction 2008"
\cite{8}.

\section{On Single Diffractive Dissociation Cross Section}

\begin{figure}[]
\begin{center}
\begin{picture}(288,192)
\put(20,10){\epsfbox{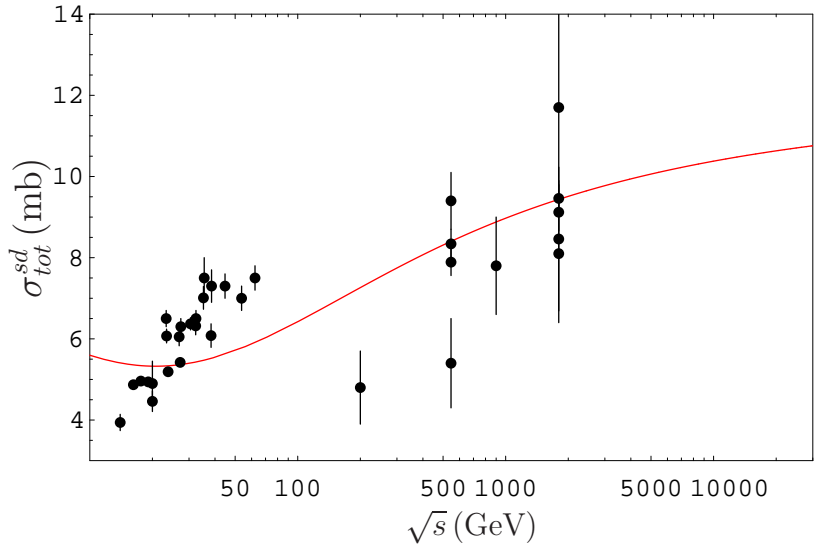}} \put(155,0){$\sqrt{s}\,
(\mbox{GeV})$} \put(5,90){\rotatebox{90}{\large$\sigma_{tot}^{sd}\,
(\mbox{mb})$}}
\end{picture}
\end{center}
\caption{Total single diffraction dissociation cross-section compared
with the theory. Solid line represents our fit to the
data.}\label{fig5}
\end{figure}

Concerning the single diffractive dissociation cross section, we have
found the bound (like Froissart bound!)
\[
\fbox{$\displaystyle\sigma_{tot}^{sd}(s) < Const,\quad s \rightarrow
\infty$}\,,
\]
and a good theoretically justified parameterization for the total
single diffractive dissociation cross section looks as follows
\cite{14}
\begin{equation}
\sigma_{tot}^{sd}(s) = \frac{A_0 +
A_2\ln^2(\sqrt{s}/\sqrt{s_0})}{R_0^2(s)},\label{sdtot}
\end{equation}
where $A_0, A_2$ are free parameters to be found from the fit to the
experimental data on $\sigma_{tot}^{sd}$. The fit yielded
\[
A_0 = 28.05\pm 0.66\, \mbox{mbGeV}^{-2},\quad A_2 = 4.99\pm 0.57\,
\mbox{mbGeV}^{-2}.
\]
The fit result is shown in Fig.~\ref{fig5} \cite{21}. As seen, the
fitting curve, as in the previous fit \cite{14}, goes excellently
over the experimental points of the CDF group at Fermilab \cite{22}.
The experimental data points for the total single diffraction
dissociation cross sections have been extracted from different papers
and collected in \cite{21}; see references therein.

One important note should be taken here. The main point of our
approach is that the fundamental three-body forces are responsible
for the dynamics of particle production processes of inclusive type.
In fact, we have found a formula expressing one-particle inclusive
cross section through the amplitude of three-body forces. Our
consideration revealed several fundamental properties of one-particle
inclusive cross-sections in the region of diffraction dissociation.
In particular, it was shown that the slope of the diffraction cone in
single diffraction dissociation process $pp\rightarrow pX$ is related
to the effective radius of three-nucleon forces in the same way as
the slope of the diffraction cone in elastic scattering process
$pp\rightarrow pp$ is related to the effective radius of two-nucleon
forces. As was demonstrated above, the effective radii of two- and
three-nucleon forces, which are the characteristics of elastic and
inelastic interactions of two nucleons, define the structure of the
total cross-sections in a simple and physically clear form.

Some time ago many high energy physicists suggested that the increase
of total cross-sections was due to the increase of single diffraction
dissociation cross sections. Now we understand that this suggestion
is wrong and, moreover, we know why this is wrong. We have
established that the phenomenon of exceedingly moderate energy
dependence of single diffraction dissociation cross-sections on $s$
discovered by CDF at Fermilab is due to effect of screening of
three-body forces by two-body ones in regime of unitarity saturation
of two- and three-nucleon forces at Fermilab Tevatron energies. In
this context, the CDF data on single diffraction dissociation cross
sections represent the significant experimental result which has to
be tested at the LHC. At the LHC we predict
\begin{equation}\label{sigma-sd}
\sigma_{tot}^{sd}(\sqrt{s}=14\,\mbox{TeV}) = 10.51\pm 1.06\,\mbox{mb}
\end{equation}
Here theoretical uncertainty in (\ref{sigma-sd}) has also been
calculated by one deviation in parameter $A_2$ in Eq. (\ref{sdtot}).

\section{Total Cross Section in Scattering from Deuteron}

Being inspired with the success in global description of the
proton-proton and proton-antiproton total cross sections we have
attempted to carry out the similar global description for the
proton-deuteron and antiproton-deuteron total cross sections.

It is well known that the total cross section in scattering from
deuteron can be expressed by the formula
\begin{equation}\label{sigmaD}
\sigma_{tot}^{\,d} = \sigma_{tot}^{\,p} +\sigma_{tot}^{\,n} -
\delta\sigma_{tot}^{\,d},
\end{equation}
where $\sigma_{tot}^{\,d}$, $\sigma_{tot}^{\,p}$,
$\sigma_{tot}^{\,n}$ are the total cross sections in the scattering
of the incident particle from the deuteron, proton, and neutron, and
$\delta\sigma_{tot}^{\,d}$ is called the defect of the total cross
section in scattering from the deuteron.

In the framework of the diffraction theory, Glauber obtained an
elegant expression for the defect of the total cross section in
scattering from the deuteron,
\begin{equation}\label{Glauber}
\delta\sigma_{tot}^{\,d} = \delta\sigma_{G}^{d} =
\frac{\sigma_{tot}^{p}\cdot \sigma_{tot}^{n}}{4\pi}<\frac{1}{r^2}>_d
\end{equation}
which is called the Glauber correction. In formula (\ref{Glauber}),
$<r^{-2}>_d$ denotes the average inverse square internucleon distance
in the deuteron.

Glauber found an attractive physical interpretation of the correction
that he obtained, and showed that it is related to configurations in
the deuteron in which one nucleon is in the shadow of another
nucleon, and describes the "eclipse" effect well known from data of
astronomical observations on decrease in luminosity of binary stars
during an eclipse. For this reason, this correction is often called
the shadowing correction or the screening effect. Moreover, it is
necessary to note the remarkable fact that formula (\ref{Glauber})
can be obtained from extremely simple, almost semiclassical
considerations presented by Glauber in the introduction to his famous
article \cite{23}.

However, it is quite clear that the Glauber correction and
diffraction formalism proposed by Glauber to derive this result may
theoretically be justified and understood only in the context of
nonrelativistic quantum mechanics. At high energies where the
processes of particle production are possible we have to apply the
relativistic quantum theory. Besides, the experiments carried out at
the accelerators have also shown that the Glauber correction
(\ref{Glauber}), although yielding a correct order of magnitude,
results in an underestimated value for the defect of the total cross
section in scattering from the deuteron.

We have considered the problem of scattering from the deuteron in a
detail using dynamical equations of the Bethe--Salpeter formalism for
a system of three particles in quantum field theory. The first, and
very important, circumstance revealed was related to the fact that
consistent consideration of the three-body problem in the framework
of relativistic quantum theory necessitates that the dynamics of the
three-particle system should include, along with pair (two-particle)
interactions, the fundamental three-particle forces as well which
cannot be expressed in terms of pair interactions. It was established
that fundamental three-particle forces are related to specific
inelastic interactions in two-particle subsystems and determine the
dynamics of special inelastic processes of interaction of two
particles known as one-particle inclusive reactions. Making quite
general assumptions, it was possible to calculate the contribution of
three-particle forces to the total cross section in scattering from
the deuteron and obtain a very simple and elegant formula with a
clear physical interpretation for the defect of the total cross
section in scattering from the deuteron. We have found that the
defect of the total cross section in scattering from the deuteron can
be represented as the sum of two terms
\begin{equation}\label{defect}
\delta\sigma_{tot}^{\,d}=\delta\sigma^{d}_{el}+\delta\sigma^{d}_{inel},
\end{equation}
where the quantity $\delta\sigma^{d}_{el}$ was called the elastic
defect, and $\delta\sigma^{d}_{inel}$ the inelastic defect. For these
quantities the following representations have been derived:
\begin{equation}\label{elineldef}
\delta\sigma^{d}_{el} = 2\, a_{el}(x_{el}^2)\,\sigma^{N}_{el}, \quad
\delta\sigma^{d}_{inel} = 2\, a_{inel}(x_{inel}^2)\,\sigma^{N}_{sd},
\end{equation}
\[
x_{el}^2=\frac{R_2^2}{R_d^2}=\frac{2B_{el}}{R_d^2},\quad
x_{inel}^2=\frac{R_0^2}{R_d^2}=\frac{2B_{sd}}{R_d^2}.
\]
It was assumed in our considerations that for both elastic and
inelastic interactions of the incident hadron with nucleons of the
deuteron, the proton and the neutron are dynamically
indistinguishable; i.e., corresponding dynamic characteristics of the
proton and the neutron are similar. For example,
$\sigma^{p}_{el}=\sigma^{n}_{el}=\sigma^{N}_{el}$,
$B_{el}^p=B_{el}^n=B_{el}$ and so on. This proposition is quite
justified if interactions occur at sufficiently high energies. It is
clear that at very low energies, it is necessary to take into account
that the proton and the neutron have different masses and electric
charge; however, formula (\ref{elineldef}) admits a natural
modification for this case.

It is natural to call the functions $a_{el}$ and $a_{inel}$ in the
right-hand side of formula (\ref{elineldef}) elastic and inelastic
deuteron structure functions, respectively. These functions have
clear physical meaning; see details in \cite{1}. It was remarkable
that we succeeded in obtaining extremely simple formulas for the
structure functions $a_{el}$ and $a_{inel}$ which have the following
form
\begin{equation}\label{a}
a_{el}(x^2)=\frac{x^2}{1+x^2}, \quad
a_{inel}(x^2)=\frac{x^2}{(1+x^2)^{\frac{3}{2}}}.
\end{equation}
Obviously, these formulas display new fundamental scaling
regularities in processes of interaction of composite nuclear
systems.

As seen, the structure functions $a_{el}$ and $a_{inel}$ have a quite
different behavior: $a_{el}$ is a monotonically increasing function
of the argument in the semiinfinite interval $0\leq x^2<\infty$, and
the domain of its values is bounded by the interval $0 \leq a_{el}
\leq 1$. The function $a_{inel}$ first increases, reaches its maximum
for $x^2=2$, and then decreases, vanishing at infinity; in this case,
the domain of its values lies in the interval $0 \leq a_{inel} \leq
2/3\sqrt{3}$. Of course, the difference between the behaviors of the
structure functions $a_{el}$ and $a_{inel}$ results in far-reaching
physical consequences. For example, at ultrahigh energies,
corresponding to $x^2\rightarrow\infty$, we find that the inelastic
defect vanishes and the elastic defect tends to two times the value
of the total elastic cross section for scattering on the nucleon,
whereas the total cross section for scattering on the deuteron
approaches two times the value of the nucleon total absorption cross
section. Therefore, at ultrahigh energies, $A$ dependence of total
cross sections for scattering on nuclei should be recovered, with the
difference that the fundamental quantity in front of $A$ is the
nucleon total absorption cross section, rather than the total cross
section for scattering on the nucleon,
\begin{equation}\label{A}
\sigma^{A}_{tot}=A\,\sigma^{N}_{inel},\quad s\rightarrow\infty.
\end{equation}

A very interesting aspect of our consideration, that the inelastic
defect in the total cross section for scattering on the deuteron is a
manifestation of fundamental three-body forces. Clearly, the Glauber
formula appears as a special case, if the inelastic defect is
neglected, and for the elastic structure function, the following
approximation (valid for $x^2<<1$) is used: $a_{el}(x^2)\simeq x^2 $,
where we have to take into account that $\sigma^{N}_{el}\simeq
(\sigma^{N}_{tot})^2/16\pi B_{el}$.
\begin{figure}[h]
\begin{center}
\begin{picture}(298,184)
\put(20,10){\epsfbox{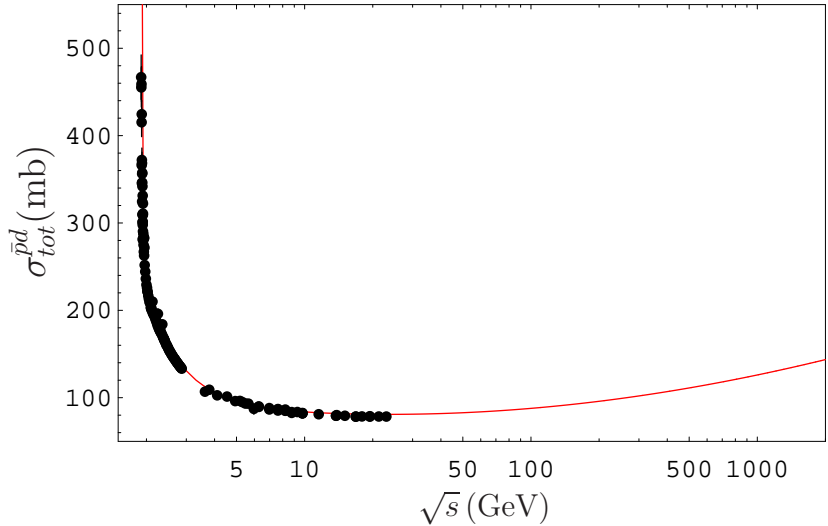}} \put(155,0){$\sqrt{s}\,
(\mbox{GeV})$}
\put(0,90){\rotatebox{90}{\large$\sigma_{tot}^{\bar{p}d}
(\mbox{mb})$}}
\end{picture}
\end{center}
\caption{The total antiproton-deuteron cross-section compared with
the theory. The experimental data points extracted from
\cite{2}.}\label{fig6}
\end{figure}

\begin{figure}[h]
\begin{center}
\begin{picture}(288,184)
\put(15,10){\epsfbox{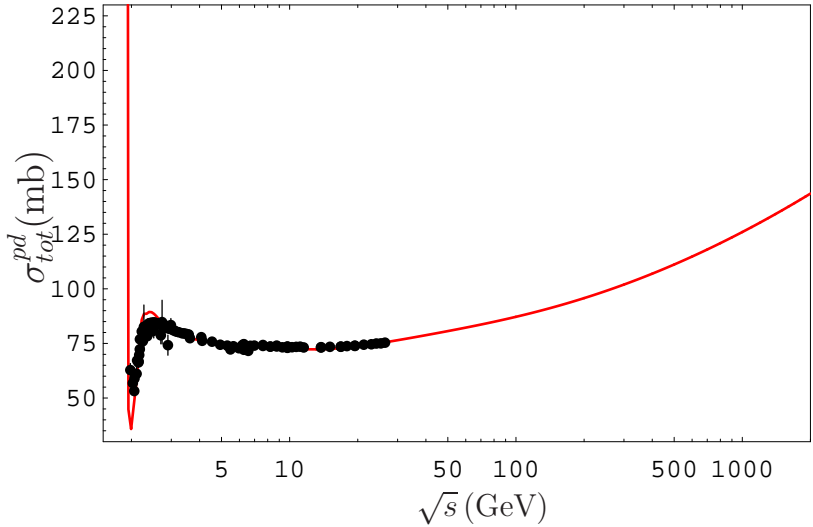}} \put(155,0){$\sqrt{s}\,
(\mbox{GeV})$} \put(0,88){\rotatebox{90}{\large$\sigma_{tot}^{pd}
(\mbox{mb})$}}
\end{picture}
\end{center}
\caption{The proton-deuteron total cross-section compared with the
theory without any free parameters. The experimental data points
extracted from \cite{2}.}\label{fig7}
\end{figure}

Of course, of special interest to us was the comparison of the
theoretical results we obtained with available experimental data on
total cross sections for the scattering of protons and antiprotons on
deuterons. Figures \ref{fig6} and \ref{fig7} show the results of this
comparison. We used the global description of $pp$ and $p\bar p$
total  cross sections and cross sections for diffraction
dissociation, taking into account the most recent experimental data
obtained by the CDF collaboration at FNAL. It should be added that
the comparison with experimental data on total cross sections for the
scattering of protons and antiprotons on deuterons was carried out in
two stages. At the first stage, theoretical calculations were
compared with experimental data on total cross sections for the
scattering of antiprotons on deuterons under the assumption that
$R_d^2$ is the only free parameter whose value should be determined
by fitting experimental data. As a result of statistical analysis,
the following value of $R_d^2$ was obtained: $R_d^2=66.61 \pm 1.16
\,\mbox{GeV}^{-2}$. The latest experimental measurements of the
deuteron matter radius indicate that $r_{d,m}=1.963(4)\,\mbox{fm}$
\cite{24}, which yields $r^2_{d,m} = 3.853\,\mbox{fm}^2 =
98.96\,\mbox{GeV}^{-2}$. The value of $R_d^2$ obtained by us
satisfies the relation $R_d^2 = 2/3\,r^2_{d,m}$. Besides, in data
analysis we have used a simplified assumption as $\sigma^{\bar p
n}_{tot}=\sigma^{\bar p p}_{tot}$. The results of theoretical
calculations are shown in Fig.~\ref{fig6} up to Tevatron energies
(FNAL). At the second stage, experimental data on total cross
sections for the scattering of protons on deuterons were compared
with theoretical calculations in which the value of $R_d^2$ was taken
as equal to that obtained at the first stage from analysis of data on
$\bar p d$  total cross sections. In other words, the curve in
Fig.~\ref{fig7} corresponds to theoretical calculations carried out
using formulas (\ref{sigmaD}), (\ref{defect}), (\ref{elineldef}) and
(\ref{a}) without any free parameter. In this figure, the results of
theoretical calculations are also shown up to Tevatron energies. As
in the previous fit we supposed $\sigma^{pn}_{tot}=\sigma^{pp}_{tot}$
and $\sigma^{pp}_{tot}$ was taken from our global description of
proton-proton total cross sections. We have also assumed that
$B^{pN}_{el}=B^{\bar p N}_{el}\equiv B_{el}$. As can be seen,
Figs.~\ref{fig6} and \ref{fig7} show quite a remarkable
correspondence of the theory to the experimental data even though the
resonance region requires a more careful consideration because
simplified assumptions we used cannot be justified at low energies.
Nevertheless, this is a remarkable fact that the dip structure of the
proton-proton total cross section at low energies manifests itself in
the proton-deuteron total cross section too.

\noindent At the LHC we predict
\begin{equation}\label{predict2}
\sigma^{\,d}_{tot}(\sqrt{s}=14\,\mbox{TeV}) = 206.86\,\mbox{mb}.
\end{equation}

\begin{figure}[]
\begin{center}
\begin{picture}(288,178)
\put(20,10){\epsfbox{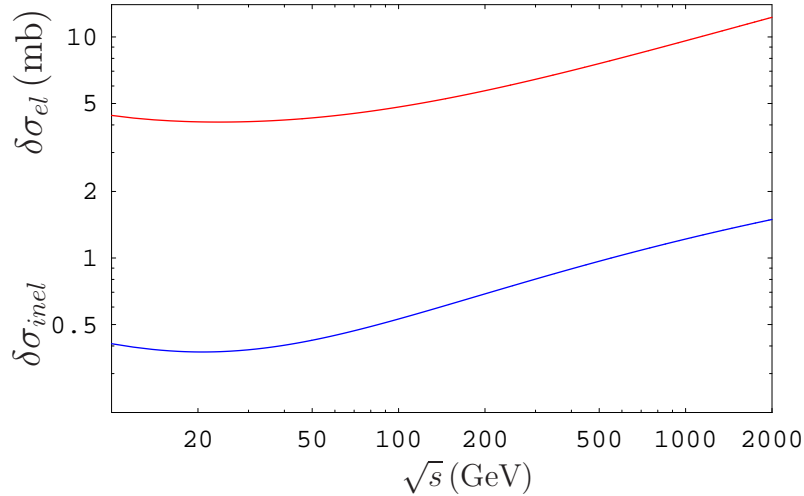}} \put(155,0){$\sqrt{s}\,
(\mbox{GeV})$}
\put(8,45){\rotatebox{90}{\large$\delta\sigma_{inel}$}}
\put(8,125){\rotatebox{90}{\large$\delta\sigma_{el} \,(\mbox{mb})$}}
\end{picture}
\end{center}
\caption{Elastic and inelastic defects of the proton-deuteron total
cross section represented by the theory.}\label{fig8}
\end{figure}

\section{On the Defect of the Total Cross Section in Scattering from Deuteron}

Figure \ref{fig8} shows the results of our theoretical calculations
of elastic and inelastic defects of the total cross section for the
scattering of (anti)protons on deuterons in the energy range 10 –-
2000 GeV. It follows from these calculations that the value of the
elastic defect is about 10\% of the total nucleon-nucleon cross
section, and the value of the inelastic defect is about 10\% of that
of elastic defect, i.e., approximately 1\% of the total
nucleon-nucleon cross section. Figuratively speaking, while the
elastic defect represents a fine structure, the inelastic defect
should be attributed to a hyperfine structure in the total cross
sections for scattering on the deuteron. In our approach, the
inelastic defect is related to the manifestation of fundamental
three-body forces; therefore, in this sense, three-body forces play
the role of "fine tuning" in the dynamics of the relativistic
three-particle system. We should give credit to experimentalists who
have created experimental setups capable of achieving the precision
of measurement that makes it possible to discriminate between
inelastic defects in total cross sections for the scattering of
particles at high energies. Along these lines, we believe that
further experimental high precision measurements of proton-deuteron
total cross sections at the LHC would also be extremely desirable.

\begin{figure}[h]
\begin{center}
\begin{picture}(288,172)
\put(20,10){\epsfbox{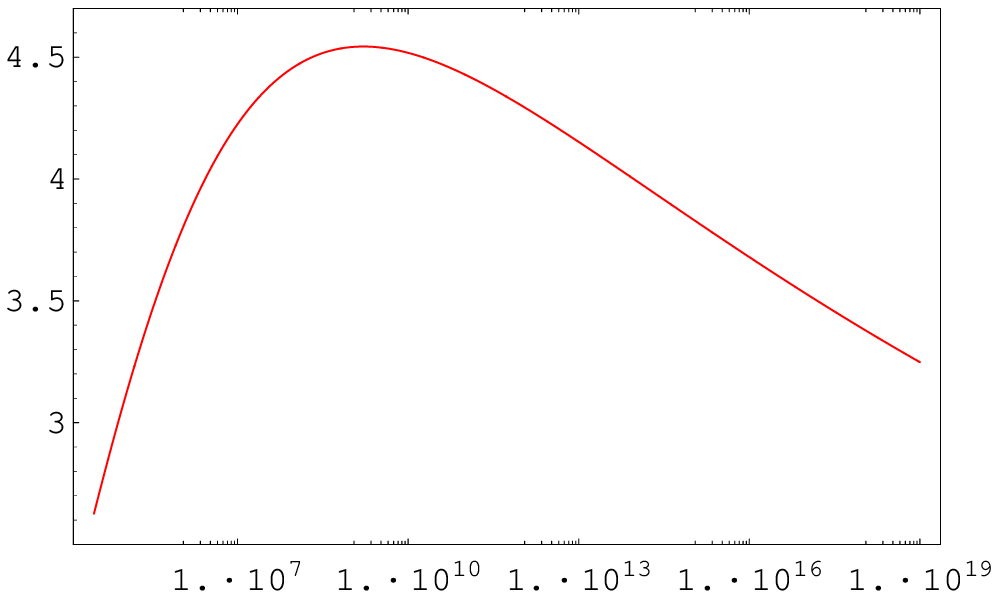}} \put(155,0){$\sqrt{s}\, (GeV)$}
\put(5,87){\rotatebox{90}{\large$\delta\sigma_{inel} (\mbox{mb})$}}
\end{picture}
\end{center}
\caption{The three-body forces contribution (inelastic screening) to
the (anti)proton-deuteron total cross-section calculated with the
theory in the range up to Planck scale.}\label{fig9}
\end{figure}

As was already noted above, the maximum value of the inelastic defect
is reached at $x_{inel}^2=2$ ($x_{inel}^2\equiv R_0^2/R_d^2$). In
other words, the energy at which the inelastic defect reaches its
maximum value is determined from the equation
$R_0^2(s_{max})=2R_d^2$. Calculations based on our analysis of
available experimental data yield $\sqrt{s_{max}}=9.01\cdot
10^8\,\mbox{GeV}=901\,\mbox{PeV}$. Obviously, such energies cannot be
achieved at either existing or designed accelerators. However,
manifestations of this effect can be sought in phenomena observed in
ultrahigh energy cosmic rays. This is the subject of a separate
investigation; see, in particular, \cite{25}. We note, however, that
$s_{max}$ has a clear physical meaning, in that it separates two
energy regions: the energy region $s<s_{max}$, in which the effective
radius of three-particle forces does not exceed the deuteron size,
or, more precisely, $1/2\,R_0^2(s)<R_d^2$, and the energy region
$s>s_{max}$, in which the effective radius of three-particle forces
becomes larger than the deuteron size, $1/2\,R_0^2(s)>R_d^2$. It
should be especially underlined that the unitarity requirement
results in the suppression of the inelastic defect at ultrahigh
energies in such a way that only the elastic part of the total defect
remains at asymptotically infinitely high energies. The existence of
$s_{max}$, at which the inelastic defect begins to be suppressed,
seems to us a very important characteristic of fundamental dynamics.
Figure~\ref{fig9} shows the inelastic defect in the region of the
maximum, calculated theoretically by us.
\vspace{1cm}

Our comparison of the theory with experimental data on total
nucleon-deuteron cross sections shows that, in the description of
particle scattering on the deuteron at high energies, it is
sufficient to take into account only nucleonic degrees of freedom in
the deuteron. A loosely bound two-nucleon system, the deuteron looks
as though the clusterization of quarks into nucleons is not
destroyed, even when nucleons come close to each other. Nucleons that
are close to each other in the deuteron do not lose their
individuality and, therefore, it is not necessary to introduce
unspecified six-quark configurations in the deuteron. The structure
for the defect of the total cross section for scattering on the
deuteron we obtained corresponds to this pattern. The general
formalism of quantum field theory admits representation of the
particle scattering dynamics on a composite system in terms of the
fundamental dynamics of particle scattering on isolated constituents
and the structure of the composite system. Probably one of the most
pleasant findings was that comparison with experimental data on
proton-deuteron and antiproton-deuteron total cross sections at high
energies already showed very good agreement of the theory with
experiment. As a matter of fact, the experimental measurement of the
proton-deuteron total cross section at the LHC might be as a crucial
test to discriminate different models for the proton-proton total
cross section proposed.

At the LHC we predict
\begin{equation}\label{predict}
\delta\sigma^{d}_{tot}(\sqrt{s}=14\,\mbox{TeV}) =
26.19\,\mbox{mb},\quad \delta\sigma^{d}_{el} = 23.88\,\mbox{mb},
\quad \delta\sigma^{d}_{inel} = 2.31\,\mbox{mb}.
\end{equation}

\section{``Soft" Physics Observables at the LHC}

Our predictions at the LHC  regarding the ``soft" physics observables
here discussed  are collected in Table \ref{tab}.

\begin{table}[h]
\begin{center}
\caption{Theoretical predictions for the ``soft" physics observables
at the LHC .}\vspace{5mm}{\scriptsize
\begin{tabular}{||c|c|c|c|c|c|c|c|c||} \hline
$\sqrt{s}\,(\mbox{TeV})$ & $\sigma^{pp}_{tot}(\mbox{mb})$ & $B_{el}(\mbox{GeV}^{-2})$ & $\sigma^{pp}_{el}(\mbox{mb})$ & $\sigma^{\,sd}_{tot}(\mbox{mb})$ & $\sigma^{\,d}_{tot}(\mbox{mb})$ & $\delta\sigma^{\,d}_{tot}(\mbox{mb})$ & $\delta\sigma^{\,d}_{el}(\mbox{mb})$ & $\delta\sigma^{\,d}_{inel}(\mbox{mb})$\\
\hline
    1.80   & 77.01  & 17.97 & 16.86 & 9.44 & 140.75 & 13.27 & 11.82 & 1.45\\ \hline
    7.00   & 101.52 & 22.21 & 23.71 & 10.22 & 182.05 & 20.99 & 18.97 & 2.02\\ \hline
    10.00  & 109.03 & 23.51 & 25.84 & 10.38 & 194.51 & 23.55 & 21.38 & 2.17\\ \hline
    14.00  & 116.53 & 24.81 & 27.97 & 10.51 & 206.86 & 26.19 & 23.88 & 2.31 \\
    \hline
\end{tabular}}\label{tab}
\end{center}
\end{table}

\section{Discussion}

The sequence of our investigations, with rather cumbersome
derivations and complicated tiresome computations, can be traced
following the references cited and references therein. Here, I would
only like to tell you a little bit about our long way to the Everest.

As a starting base the Quantum Field Theory with local fields
satisfying the standard set of axioms has been chosen. We have
builded, in the first, the constructive Bethe-Salpeter formalism in
any $n$-particle sector ($n\geq 2$) without having the use of
perturbation theory. In fact, we have restricted to two-particle and
three-particle sectors in a detail. It turned out in relativistic
quantum theory the dynamics of three-particle system contained with a
necessity the fundamental three-body forces. Actually, the
fundamental three-body forces take place in any multiparticle system
where the number of particles is greater than two. The three-body
forces represent the defect of total three-particle interaction over
the sum of two-body forces and describe the true three-body
interactions. Three-body forces are an inherent connected part of
total three-particle interaction which cannot be reduced to the sum
of pair interactions.

The Great Froissart Theorem was as a Guiding Star in our studies. The
Froissart bound for the total cross section of two-body reaction
$a+b\rightarrow a+b$ can be written in the form \cite{26}
\begin{equation}
\sigma_{ab}^{tot}(s)<4\pi R_2^2(s).\label{Froissart}
\end{equation}
Here the quantity $R_2(s)$ has a strong mathematical definition with
a clear and transparent physical meaning; see details in Ref.
\cite{26} and references therein
\begin{eqnarray}
R_2(s)&\stackrel{def}{=}&\frac{L}{|\bf q|} =
\frac{2\sqrt{s}\ln{\tilde
P}_2(s)}{\sqrt{2\epsilon(s)\lambda(s,m_a^2,m_b^2)}} = \frac{\ln
{\tilde P}_2(s)}{\sqrt{t_0}}\\
&\simeq& \frac{9}{4\sqrt{t_0}}\ln (s/s_0) = \frac{9}{8m_\pi}\ln
(s/s_0),\quad s \gg s_0, \quad (t_0\equiv 4m_{\pi}^2).
\label{effectiveradius}
\end{eqnarray}
The pion mass $m_{\pi}$ in R.H.S. of Eq. (\ref{effectiveradius})
appears from the nearest $t$-channel threshold, $s_0$ is a
determinative scale usually extracted from a fit to experimental
data. The quantity $R_2(s)$ is named as the effective radius of
two-body forces, and it is simply related with the experimentally
measurable quantity which is the slope of diffraction cone
$B_{el}(s)$ in elastic forward scattering for the two-body reaction
\begin{equation}\label{B-el}
B_{el}(s) = \frac{1}{2}R_2^2(s).
\end{equation}
It should especially be emphasized that the quantity $R_2(s)$
accumulates all information concerning polynomial boundedness  and
analyticity of the two-body reaction amplitude in a topological
product of complex $s$-plane with the cuts ($s_{thr}\leq s
\leq\infty,\, u_{thr}\leq u \leq\infty$) except for possible fixed
poles and circle $|t|\leq t_0$ in complex $t$-plane, where $s$, $t$,
$u$ are Mandelstam variables. That analyticity is proved in the
framework of axiomatic Quantum Field Theory, and this is enough to
save and extend the fundamental Froissart result previously obtained
at a more restricted Mandelstam analyticity in the framework of the
analytic $S$-matrix theory. The corner stone in that extension has to
be referred to Harry Lehmann \cite{27} who proved that two-body
elastic scattering amplitude is analytic function of $\cos\theta$,
regular inside an ellipse in complex $\cos\theta$-plane with center
at the origin. The fundamental Jost-Lehmann-Dyson representation --
brilliant quintessence of general principles in the theory of
quantized fields -- especially Dyson's theorem for a representation
of causal commutators in local Quantum Field Theory \cite{28,29,30}
and not more have been used by Harry Lehmann. From the fundamental
result of Harry Lehmann it follows that the partial wave expansions
which define physical scattering amplitudes continue to converge for
complex values of the scattering angle, and define uniquely the
amplitudes appearing in the nonphysical region of non-forward
dispersion relations. In fact, expansions converge for all values of
momentum transfer for which dispersion relations have been proved.

The Froissart bound represents a physically tangible consequence from
abstract mathematical structures given by general axioms in the
theory of quantized fields. That is why, the Froissart bound is often
considered as intrinsic property of the theory of quantized fields.

In our opinion, the bound (\ref{Froissart}) represents the most
rigorous mathematical formulation of the holographic principle
\cite{31} which is widely discussed in the recent literature. Thus
the holographic principle has been incorporated in the general scheme
of axiomatic Quantum Field Theory and resulted from the general
principles of the theory of quantized fields \cite{26}.

From the Froissart bound in the case of the two-body forces saturated
unitarity one obtains
\begin{equation}\label{unitaritysaturation}
\sigma_{ab}^{tot}(s) = 4\pi R_2^2(s)\simeq C_{ab}\ln^2(s/s_0), \quad
s\rightarrow\infty,
\end{equation}
where
\begin{equation}\label{constant}
C_{ab}=\frac{4\pi\cdot 81}{64\,m_{\pi}^2}=\frac{15.9}{m_{\pi}^2}\cong
339\, \mbox{mb}.
\end{equation}
Certainly, the value 339 mb for the constant $C_{ab}$ is too large to
fit to available experimental data, and this is a week place of the
general theory. However, we cannot exclude that the two-body forces
may not saturate unitarity in the range of reachable energies at now
working accelerators. On the other hand, it is quite clear that Eq.
(\ref{Froissart}) really represents the bound only, and we have to
find the physical arguments to compare the general theory with
experiment. Actually, we have found an elegant way for
structurization of the constant $C_{ab}$ in R.H.S. of Eq.
(\ref{unitaritysaturation}) if we have taken into account not only
two-particle but three-particle unitarity as well.

The Froissart bound in any $n$-particle sector ($n\geq 2$) can be
written in the following form:
\begin{equation}
Im\, {\cal F}_n(s;\cos\omega=1) <
J_n(s)S_{D-1}[R_n(s)]^{D-1},\label{FroissartG}
\end{equation}
where ${\cal F}_n(s;\cos\omega)$ is the amplitude of $n$-body forces,
$\cos\omega={\bf e}'\cdot{\bf e}$, ${\bf e}$ and ${\bf e}'$ are two
unit vectors on $(D-1)$-dimensional sphere $S_{D-1}$, which
characterize the initial and final states of $n$-particle system,
dimensionality $D$ of multidimensional space is related to the number
of particles $n$ by the equation $D=3n-3$, $J_n(s)\sim s^{n/2}$ is
$n$-particle flux, $S_{D-1}=2\pi^{D/2}/\Gamma(D/2)$ is a surface of
$(D-1)$-dimensional unit sphere,  $R_n(s)\sim \ln(s/s_0^{'})$ is the
effective radius of $n$-body forces; see details in \cite{26}.

As mentioned above, the structure given by Eq.~(\ref{sigmatot}) has
been appeared as consistency condition to fulfil the unitarity
requirements in two-particle and three-particle sectors
simultaneously. It is a non-trivial fact that the constant in R.H.S.
of Eq. (\ref{asigma}), staying in front of effective radius of
two-nucleon forces, is 4 times smaller than the constant in the
Froissart bound. But this is too small to correspond to the
experimental data if we use the experimental data on $B_{el}(s)$. The
second term in R.H.S. of Eq. (\ref{asigma}) fills an emerged gap.

It is interesting to note that in the case of the two-body forces
saturated the Froissart bound, taking into account that
$\sigma_{el}=\sigma_{tot}^2/16\pi B_{el}\,(\rho_{el} = 0)$, and
$B_{el}=R_2^2/2$, one obtains
\begin{equation}\label{Pumplin}
\sigma_{ab}^{tot}(s) = 4\pi R_2^2(s)\quad \Rightarrow \quad
\sigma_{ab}^{el}(s)=\frac{1}{2}\,\sigma_{ab}^{tot}(s),\quad
s\rightarrow\infty.
\end{equation}
Thus we come to the following statement: {\it The two-body forces
saturated the Froissart bound saturate the Pumplin bound as well}.

Of course, following the general scheme of the local quantum field
theory, we must not forget about the crossing
\[
\sigma_{tot}^{p\bar p}(s) = \sigma^{(+)}(s) + \sigma^{(-)}(s), \qquad
\sigma_{tot}^{pp}(s) = \sigma^{(+)}(s) - \sigma^{(-)}(s),
\]

\[
\sigma^{(+)}(s) = \frac{1}{2} \{\sigma_{tot}^{p\bar
p}(s)+\sigma_{tot}^{pp}(s)\},\qquad \sigma^{(-)}(s) =
\frac{1}{2}\{\sigma_{tot}^{p\bar p}(s) - \sigma_{tot}^{pp}(s)\}.
\]

\section{Conclusion}

As was demonstrated above, the effective radii of two- and three-body
forces being the characteristics of elastic and inelastic
interactions in two-body subsystems have been combined in a special
form determining the nontrivial dynamical structure for the total
cross section clearly confirmed by the experimental data. The further
experimental confirmation of this dynamical structure for the total
cross section at the CERN Large Hadron Collider is a good task.

It would be very important to experimentally investigate the ``soft"
physics by the measurements of all above mentioned observables
simultaneously at one and the same device which the CERN Large Hadron
Collider is.

We believe that further experimental high precision measurements of
proton-deuteron (in general proton-nucleus) total cross sections at
the LHC would also be extremely desirable.

\section*{Acknowledgements}

It is my great pleasure to express thanks to the Organizing Committee
for the kind invitation to attend this beautiful Conference. I would
like to especially thank the local organizers for warm and kind
hospitality throughout the Conference.

\end{document}